\documentstyle[onecolumn,referee,psfig]{mn}

\title[ Dark Halo Density Profile in Galaxies  ]
{\bf THE  CONSTANT DENSITY REGION OF   THE DARK  HALOS OF SPIRAL GALAXIES }  \author[
P.Salucci ] {Paolo Salucci$^{1}$ \\
$^1$SISSA -- International School for Advanced Studies, Via Beirut 2, Trieste, I-34014, 
salucci@sissa.it}

\begin{document}
 
\maketitle

\def\mincir{\raise -2.truept\hbox{\rlap{\hbox{$\sim$}}\raise5.truept
\hbox{$<$}\ }}
\def  \magcir{\raise -2.truept\hbox{\rlap{\hbox{$\sim$}}\raise5.truept
\hbox{$>$}\ }}
\def\ref{\par\noindent\hangindent 20pt}
\def \dvd{\nabla_d}
\def \dvh{\nabla_h}
\def \dv{\delta V}
\def \dvlum{\delta V_{lum}}
\def \Ro{R_{opt}}
     
\begin{abstract}
We  determine a crucial feature of  the dark halo density distribution
from the fact that the luminous matter  
dominates  the  gravitational potential  at  about one disk scale-length
$R_D$, while at the  optical edge  $R_{opt}\simeq 3 R_d$  the   
dark matter has become the main component of  the  galaxy density.
From the kinematics of  137 spirals  we find  that  the  
DM halo density profiles  are
self-similar at least  out to   $R_{opt}$  and show  core radii  
much larger than the corresponding  disk scale-lengths. The luminous 
regions of  spirals consist of stellar  disks  
embedded in  dark halos with  roughly constant density. 
This   {\it invariant}  DM  profile  is very  difficult to reconcile 
with the fundamental properties of the   density distribution of  
CDM halos. With respect to previous work, the present evidence  
is  obtained  by means of a  {\it robust}  
method and    for a {\it  large}   and {\it complete} sample of {\it normal} spirals. 
\end{abstract}

\section{Introduction}

Rotation curves (hereafter RCs) of spiral  galaxies do
not show any Keplerian fall-off: this implies the presence of an invisible mass
component (Rubin et al. 1980; Bosma 1981; see Salucci and Persic,
1997). More precisely, the mass {\it distribution} of stars and gas does not
match that  of the gravitating matter (Persic and Salucci  1988, 
Persic, Salucci and Stel, 1996, hereafter PSS,  
see also Corbelli and Salucci, 1999);  the  discrepancy increases 
with increasing radius and, 
at a given  radius measured  in units of disk length-scales $R_d$,  
it  increases with decreasing galaxy luminosity 
(Persic \& Salucci 1988, 1990a,b; Broeils 1992). 

Each {\it individual} circular velocity  $V(R)$ and the spiral  {\it
Universal Rotation Curve}  can both be
 represented, in  the optical regions,  
by a  linear law:  (Rubin et al. 1980, see  Persic \& Salucci (1991)  and  PSS for details)
$$ 
V(R) ~ \simeq V_{opt}\Big[1+ \nabla ~ \Big({R\over{R_{opt}}}-1\Big)\Big] ~~~~~~~
~~~~~~~~~ 0.4 \mincir R/R_{opt} \mincir 1.2 ~~
\eqno(1a)
$$

\noindent  where $R_{opt}=3.2 R_d$, $V_{opt}\equiv V(R_{opt})$, and 
$$
\nabla(V_{opt})=0.10-1.35 \Big( log {V_{opt}\over{200\ km/s}} \Big)
\eqno(1b)
$$
In (1b) the range is  
$-0.1 \leq  \nabla\equiv {d{\rm log} V(R) \over d {\rm 
log} R} \biggr|_{R_{opt}} \leq 0.7$ and    
the   r.m.s. is  0.05 (see PSS and Fig 1)).
\footnote{We  take  $R_{opt}$ as the reference 
disk scale to follow PSS. We can  
use any other  multiple of  $R_d$  to specify  the URC: no result of this 
paper changes.  The URC for the whole range
of available data,  $0.1 R_{opt}\leq R \leq 2 R_{opt}$, is given in PSS}
Notice that we will freely interchange
luminosities and  $V_{opt}\equiv V(R_{opt})$, given their high degree of correlation.    

The analysis  of the URC and/or   of individual RC   has  provided crucial 
knoweledge on  the main 
{\it global} properties  of the dark and  luminous matter
(e. g.  Salucci \& Persic, 1997). 
On the other hand,  to invesatigate  the local properties of  
the dark halos  (e.g.  the central density)  is   quite difficult,  
especially for    luminous matter  (LM) dominated objects 
(van Albada et al 1985, but  see the 
exception  of   high-resolution RC's,  Borriello and
Salucci, 2000).  
However, as a result of recent substantial observational and theoretical  
progress, a  proper  investigation of  the spiral's halos  
mass structure is now possible.
In fact,  from the study of a
  large number of  high-quality  RC's recently   
available (Persic \&
Salucci, 1995) it  has emerged that:  

\noindent {\it i)}  the dark matter  follows a   regime of 
Inner Baryon Dominance (IBD) according
to which   in  every  normal spiral  there is 
a {\it transition}   radius  $R_{IBD}$
$$
R_{IBD}\leq 2 R_d  \Big({V_{opt}\over{200 km/s}}\Big)^{1.2}
\eqno(2)
$$
inside which  the luminous matter  totally  
accounts for the whole mass distribution (see Ratnam and Salucci 2000, 
Salucci and Persic 1999ab,  
Salucci et al, 2000). This allows us to address the  issue  of 
the  degeneracy problem raised  by Van Albada, et al. (1985): 
the URC (and  individual rotation curves) 
{\it do} show, in their profile,   the
kinematical signature  of a transition between an  inner LM-dominated 
region and  an outer DM-dominated one (e.g. Salucci and Persic, 1999b).

\noindent {\it ii)}  dark halos  are  distributed very differently 
from the various    "luminous" components (Corbelli and Salucci, 1999).

These  findings allow us  1) to improve  the determination of the    disk mass  
to the  level required  
for  investigating  the  DM distribution and  2)  
to relate  the   dark halos   around galaxies to   
collision-less  non-baryonic  cosmological structures.

The aim  of this letter is  to derive, for a  large and 
complete sample of  spirals,  
the     density  profile  of  the dark  halos  at  the  edge 
of the disks which are embedded in them and reveal  dark constant-density  
regions  of  obvious  cosmological  importance. 
The  evidence   for  core radii
of   DM  halos,   obtained so far  by means of   mass modeling 
of a (small) number of  DM-dominated RC's (e.g. Flores and Primack 1994, 
Moore 1994, Burkert 1995),  
is  being   questioned in the light 
of the intrinsic   uncertainty of the analysis itself
(Burkert and Silk, 1997, van den Bosch 1999). In many cases, in fact,
the standard RC  fitting method has difficulty in    discriminating    
a NFW (Navarro, Frenk and White, 1996)  density  profile,
 which has    $V_h(R)\propto R^{1/2}$ in the center, from  a 
constant-density one, with  $V_h(R) \propto R$.

We  tackle this  issue by  resorting to the  method  of Persic and 
Salucci (1990b) in which 
the power law slope of the dark halo 
velocity at the disk edge is derived by means of  a  robust and 
straightforward procedure, 
which ultimately exploits the fact that,  at $\sim R_{opt}$,  the dark  halo  
is  always  the   main  {\it density} component,  even  when it is a 
negligible {\it mass} component at about $R_D$.  
\footnote {This can be  seen by combining  
eq (1) with the Poisson equation (see, e.g.      
Fall and Efstasthiou,  1980).}

In detail, the   goal of this letter is to detect  a clear and reliable  
feature  of the dark matter distribution, relevant  on  its own and  
(probably)  at variance with  the structural  properties of  {\it
standard}  CDM  halos. Let us stress that the issue  of  
establishing  the actual   CDM halo properties,  or  that   
of  investigating  whether non-standard  CDM scenarios may be  in 
agreement  with observations, are beyond the scope of this work.     

 \begin{figure}
\vspace{-0.6cm}
\centerline{\psfig{figure=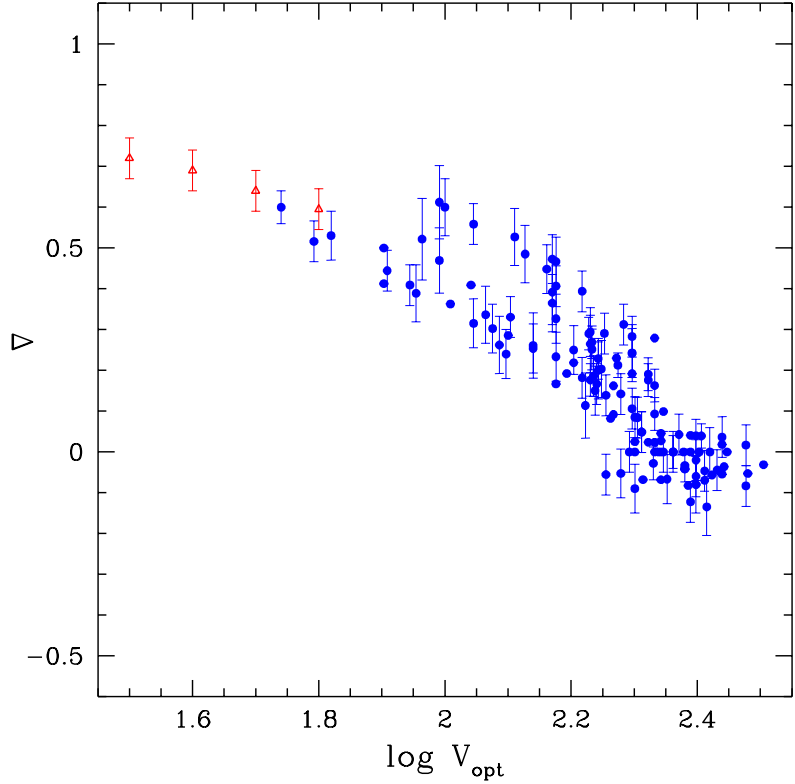,height=11cm,width=11cm}}
 
\caption{Rotation curve slope at $R_{opt}$ as a function of $V_{opt}$. Triangles refer to dwarf spirals.}
\end{figure}
 The plan of this letter  is the following: in section 2 we describe the RC  sample
and we derive the local DM  slope   $\nabla_h$ for all of  the objects, 
 in section 3 we compare the observed  halo mass distributions with the CDM prediction and  comment on our main results.

\section{The halo mass distribution at $R_{opt}$}

 Let us start 
 from the condition of rotational  equilibrium: 
$$
V^2(R) ~=~ V^2_{lum}(R)~+~V^2_h(R)\,,
\eqno(3a)
$$
where $V^2_{lum}(R)=V^2_{g}(R)+V^2_d(R)+V^2_{b}(R)$, with obvious notation, is   
  the quadratic sum of the three "luminous" components:
gas, disk, and bulge. We  define  $\beta \equiv { V^2_{lum}(R) \over V^2(R) } \biggr|_{R_{opt}}$
and  $ \beta_d \equiv { V^2_{d}(R) \over V^2(R) } \biggr|_{R_{opt}}$ .
 In late type spirals,  the  exponential thin  
disk (of mass   $M_d$)  
is by far the main contributor to $V_{lum}(R)$, at   $R \simeq R_{opt}$  (e.g. Verheijen, 1997;
Rhee, 1997;  Persic, Salucci \& Ashman,  1993), and   so 
 $$
\beta\simeq \beta_d
\eqno(3b)
$$ 
with :   
$$
V^2_{d}(x)=\beta_d V_{opt}^2 x^2{(I_0K_0-I_1K_1)|_{1.6 x}\over{(I_0K_0-I_1K_1)|_{1.6 }}}
\eqno(4)
$$ 
 where $x=R/R_{opt}$, $I_n,K_n$ are the modified
 Bessel functions of $nth$-order, and  using the  previous definitions,
 $V^2_h(R_{opt})$,  the DM velocity contribution at $R_{opt}$ is given by 
$$
  V^2_h(R_{opt})=(1-\beta) V^2_{opt}
\eqno(5)
$$
 
 Let us define: 
$$
\nabla_{d} ~\equiv ~{d{\rm log} V_{d}(R) \over d {\rm log} R} \biggr|_{R_{opt}}\, 
\eqno(6a)
$$ 
$$
\nabla_{lum} ~\equiv ~{d{\rm log} V_{lum}(R) \over d {\rm log} R} \biggr|_{R_{opt}}\, 
\eqno(6b)
$$ 
and     $\nabla_h$, $\nabla_{g}$,  $\nabla_b$  in the 
same way of the l.h.s of equation (6a)  when   $V_d$ is  substituted by  
$V_{h}$, $V_g$ and  $V_b$ (the subscripts $h, g, b$ refer to halo, gas and 
bulge respectively).  
In view of  the argument {\it ii)} in  the previous section,
we can  identify  $V_h(R)$ with  the contribution of  a    non-baryonic dark component. 

From the previous eqs. 
$$
\nabla_{lum} =\Big({V_d(R_{opt}) \over{ V_{lum}(R_{opt})}}\Big)^2\nabla_d + 
\Big({V_g(R_{opt}) \over{ V_{lum}(R_{opt})}}\Big)^2 \nabla_g+ \Big({V_b(R_{opt}) \over{V_{lum}(R_{opt})}}\Big)^2 \nabla_b
\eqno(6c)
$$ 
For a bulge-less  gas-free spiral, 
 $$
\nabla_{lum}=\nabla_d=-0.273
\eqno(7a)
$$
from eqs. (4) and (6a). 
In this case,  given its definition and the self-similarity of spiral  disks,
$\nabla_{lum}$   is strictly a 
constant. The contribution of  bulge to 
 $\nabla_{lum} $ can be totally neglected  (see  Persic, Salucci and 
Ashman, (1993) for details) while that  of the gaseus  disk
can be evaluated in spirals with HI measurements: in a large sample (Rhee, 1997) we find  
that typically  
$({V_g(R_{opt}) \over{ V_{lum}(R_{opt})}})^2\simeq 3-6\times 10^{-2}$,
 and $\nabla_g \simeq 0.5 $  that leads to  $\nabla_{lum}\simeq 0.9 \nabla_d$. 
For a sub-sample of the present sample we have  (PSS):
 $$
\nabla_{lum} = -0.24 \pm 0.03 (2\sigma)
\eqno(7b)
$$
 It is worth to stress that
no result of this paper changes  1)by   assuming $\nabla_{lum}$ according to eq
(7a) rather than to eq (7b),  or  2)by  neglecting the
 (small)  variance of $\nabla_{lum}$ among spirals.

By differentiating eq. (3a)   we arrive to
(Persic \& Salucci 1990b):
$$
\nabla_h ={-\beta  \nabla_{lum} + \nabla\over{  1-\beta}}
\eqno(8)
$$
which expresses the DM halo velocity  slope in terms of the RC slope and  of the
LM  mass fraction at  $R_{opt}$.

Note  that,  for   
high-quality RC's ($\delta \nabla< 0.05$),  
the estimate of $\nabla_h$ from  eq (8) is very 
robust. In fact,   
in  DM dominated objects, ($\beta<0.5$), the  
uncertainties on  $\beta$  do not affect      
the estimate of  
$\nabla_h$,  while in  LM-dominated objects,  ($\beta>0.5$),
a reasonably  good knowledge of   $\beta$: $\delta \beta<  0.1$ suffices to   estimate
  $\nabla_h$ within a reasonable   uncertainty (i.e.     
$\delta \nabla_h<0.2$).

Once  we include    the radial dependences  of all the  
quantities, eq (8)   is in principle  valid   at any radius. However, 
for $R<R_{opt}$,  $\beta(R) \rightarrow 1$ very quickly,  so that  
even  a small    error
in $\beta$ or  $\nabla$  will  strongly  affect the estimate of  $\nabla_h(R)$. 
On the other  hand, for $R> R_{opt}$,  the number of available 
RC's rapidly decreases.    

\begin{figure}
\centerline{\psfig{figure=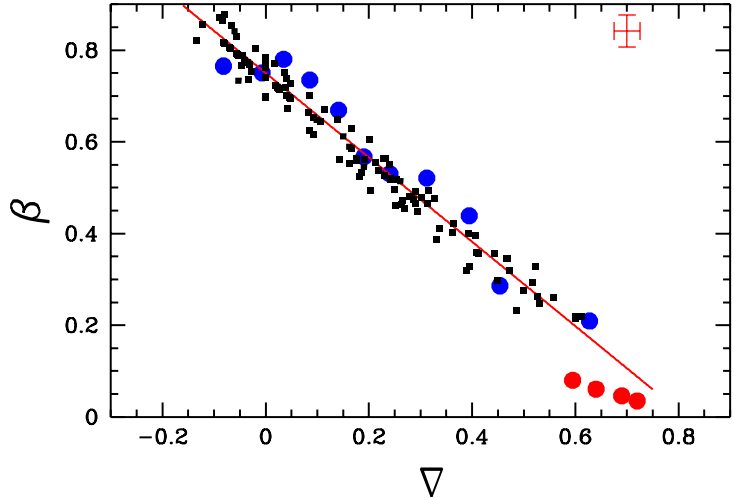,height=15cm }}
\vspace{-7cm}
\caption{The $\beta$ {\it vs }$\nabla$ relationship for the present sample.
(small filled circles) with their best fit (solid line). 
For a comparison we plot also  the  values derived from the URC and binned
over $V_{opt}$ as in PSS (large
filled cirles).}
\end{figure}

The   sample of RC's  we  use
in this letter  is presented in PSS and Salucci and Persic (1997). It
includes 131  rotation curves  of spirals and  6  for  dwarfs,  all with 
a  reliable profile
out to $R_{opt}$. This has been ensured by  the following selection 
criteria: each  RC  {\it (a)} extends out to a  $R\geq R_{opt}$;
{\it (b)} has at least 30 velocity measurements distributed homogeneously with
radius and between the two arms; and {\it (c)} shows no asymmetries or
 non-circular motions. For 21-cm RCs we  require that 
the beam-size should be  $\leq 1/2  R_d$ in order  to limit the 
uncertainties due to beam smearing. Incidentally, no LSB RC is found to  
satisfy these conditions, and this points to an intrinsic  difficulty 
in studying these objects. 

The values of  the  $\nabla's$  are   estimated by fitting with eq (1)  each of 
the 137 RC's of the
sample:   the related  uncertainty is  small: 
 $\delta \nabla \sim 0.02-0.05$;   
the values of $V_{opt}$,  $\nabla$ and  $\delta \nabla$,
are given in Tables 1 and 3 of  PSS and shown  in  Fig (1) as 
 function of $V_{opt}$.  
It is worth to point out that  {\it a)}  the range of $\nabla$ {\it along} the
luminosity  sequence is  large,  $\sim 0.7$,
while, the corresponding r.m.s., {\it at a  given} luminosity is  small  $<0.1$: the 
DM properties are likely to vary with galaxy global properties rather than random,  
{\it b)} we do not assume the URC (i.e. eq. (1b)): the values of $\nabla$ are
derived  from  each RC  .

The  LM fraction $\beta$ is  derived as   in Salucci and Persic  (1999a),  
Salucci et al. (1999) and  Ratnam \& Salucci  (2000),  i.e. by  fitting  
the inner parts of each   rotation curve,  ( 
$0\leq R\leq R_{IBD}$), with (only)  the  circular velocity of  
an exponential thin disk   given by Eq. (4). 
\footnote{In very few cases a bulge component has been added}. In this  
region, the only-disk  mass model reproduces,
with no free parameters, the {\it normalized}  rotation curve $V(R)/V_{opt}$ 
and, with a suitable choice of the  parameter
$\beta$,  the full curve  $V(R)$. This is shown for coadded RC's
in Figure (3)  and for individual RC's in  Salucci et al., (2000)
 and   in 
Ratnam and Salucci (2000)).  
The   excellent match we  always find   leads to  
very precise determinations   of  $\beta$'s; in fact, 
the $1-\sigma$ fitting uncertainties are quite modest,   $\delta \beta \simeq
0.05-0.03\beta $.
Let us  notice that the 
present method    does not   assume  a   "maximum disk" solution,  
rather,  it is   intimely related with the idea   that the  mass  distribution in  spirals
follows the
 regime of   Inner Baryon Dominance proposed    by
  Salucci \& Persic, (1999b) and then  supported by Ratnam \& Salucci, 2000 and
Salucci et al, 2000.  
Finally,  the present  method computes
$\nabla_h$ independently of  the value of the  disk  mass-to-light ratio whose   
uncertainty,  \footnote{Since  $M_d=1.1 G^{-1} \beta_d V_{opt}^2 R_{opt}$
this is     $ \sim (\beta +\delta \beta)/ (\beta -\delta \beta) $},
large  in
DM-dominated objects,  is  therefore irrelevant for  the aim of this work.

The disk mass fraction $\beta$  correlates  tightly with  $\nabla$ (see Fig (2)).
In addition,  at a fixed $V_{opt}$,   part  of  its    scatter 
originates from observational  errors and then it  is  unrelated to a    
cosmic variance of the halo  mass  
distribution. In any case, conservatively,  we  derive  the   r.m.s. of
the above relation 
by performing  the usual  least squares fit
$$
\beta= 0.75-0.95 \nabla \ \ \ \ \ \ \ \ \ \ \ \ \ \ \ \ \ \ \ \ \ \  r.m.s.=0.05
\eqno(9)
$$
A relationship like    eq.  (9)  has    been found  from  the   mass modelling of
(smaller samples of) spirals  (e.g.  Persic and Salucci 1988;  1990a; 1990b); it can
be considered   as  the basic  law  of   the  dark-luminous 
coupling in galaxies (see Salucci, 1997). 

We  derive the values of  $\nabla_h$ for the objects
of our sample by  setting $\nabla_{lum}=-0.24$ and
inserting in eq (8) the
corresponding   values  of $\beta$ and $\nabla$. 
In Fig (4) they are plotted   as a function of $ log V_{opt}$: 
we immediately realize that  $\nabla_h$   is roughly   constant over the 
whole sample and it shows no systematic  variations  along  the 
luminosity sequence. Such variations, if present,  should have clearly  appeared 
given the high-precision measurement  of $\nabla_h$. In fact 
the uncertainties $\delta \beta< 0.04$ on $\beta$ propagate into eq (8) in a modest
way, as it can be realized by 
differentiating eq (8) and combining it with eq (9) to get 
$
\delta \nabla_h \simeq (0.25+\nabla)^{-1} \delta\beta<0.1$

The average  value found for  $<\nabla_h>=0.8\pm 0.06$ indicates  that, 
at $R_{opt}$, the halo RC is  steeply increasing, marginally compatible 
with a solid body law,  $V_h\propto R$.  Consequently, 
around  $R_{opt}$, the halo density must decrease with radius 
less steeply  than 
$R^{-0.4}$. By assuming  a pseudo-isothermal density distribution,
$\rho_h(R)\propto (R^2+a^2)^{-1}$, it follows that $a> 1.3 R_{opt}$, 
i.e. a core radius significantly  larger than $R_d$,  the size 
of region where the bulk of the stellar component  is located.

We stress the  robustness of this result by noticing that,  from eq (8),
crucially  lower  values of  $\nabla_h$  (i.e. $\nabla_h<0.5$)  are 
possible   only  in the particular
combination  of  a  "flat" RC,  (i.e. for  $\nabla< 0.2$)  {\it and} a "light disk" (i. e.   
 $\beta< (0.5-\nabla)/0.77\equiv \beta_{0.5}$). Also this extreme case, 
however,  must be ruled out since
such a flat  RC cannot be fitted  by  a  mass model with 
($\beta$, $\nabla_h$)  less than or equal to ($\beta_{0.5}$, $0.5$).
(Salucci and Persic, 1999b).

\begin {figure}
\vspace{-6cm}
\centerline{\psfig{figure=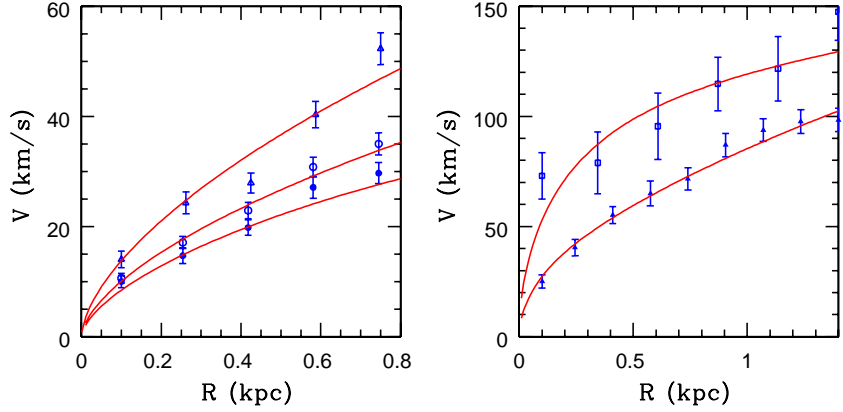,height=12cm,width=12cm}}
\caption{The innermost regions of  Universal Rotation Curve of galaxies of different luminosity
(points) compared with the disk (+bulge)  models (solid lines). Details are in Salucci et. al,
(2000)} 
\end{figure}

\section{Discussion}

We have investigated the mass distribution of DM halos for a 
large number  of spirals  in a way which is     complementary 
to the mass modeling of DM-dominated rotation curves.
This method has   been   applied  to  RC's of galaxies of all  
luminosities,  including the most  luminous ones
for which the standard  mass modeling  is quite   uncertain.    
In detail,  for each halo  we have  derived   a  single but most 
crucial structural 
property, namely,  its velocity slope  at $R_{opt}$. 

The results are impressive:  the halo mass profiles at $R_{opt}$  turn out to 
be {\it i)}  independent of the galaxy  properties,     
 {\it ii)}  Universal and  {\it iii)}  essentially featureless in the 
sense that for  any spiral the stellar disk is embedded within a 
constant density sphere. 
 
  \begin{figure}
  \centerline{\psfig{figure=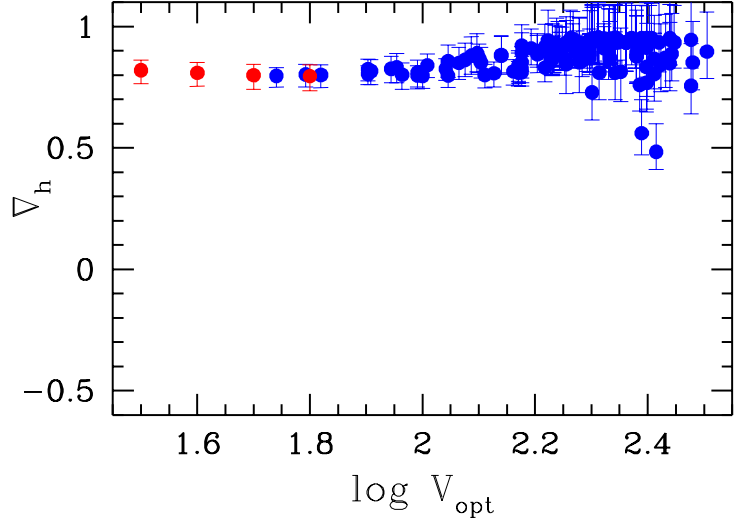,height=15truecm,width=15truecm}}
  \vspace{-7cm}
  \caption{The dark halo slopes $\nabla_h$ as a function of $V_{opt}$. 
  As a comparison, in CDM  $-0.1<\nabla_h\leq 0.5$ }
  \end{figure}

 High-resolution  simulations and analytical studies of the 
 Cold Dark Matter scenario have  pointed to a
universal  halo  density profile 
 (Navarro, Frenk and  White, 1996)
 $\rho_{CDM}(R)  \propto  R^{-1} (R+R_s)^{-2}$, with  $R_s$ being a function of the dark 
halo mass $M_{200}$,  of   the assumed 
cosmological  parameters,  
and of the  red-shift of formation of the halo. Moreover,  halos with 
identical values of  the above   quantities,  can   still  have  different 
$R_s$ if  formed at various  red-shifts and/or
assembled through   different   merging histories. In detail,  
the circular velocity  of a  CDM halo is given by:
$$
V_{CDM}(x)^2 \propto x^{-1} (-cx/(1+cx)+ln(1+c x)  ) \ \ \ \  \ 4\mincir c(\Omega_0, M_{200})<30
\eqno(8)
$$
where  $c$, is the concentration parameter,  $x\equiv R/R_{200}$,  and  $R_{200}(M_{200},z)$ is the
halo virial radius defined by $4/3\pi R_{200}^3 200 \rho_c \Omega_0 (1+z)^3 =M_{200}$.

Before proceeding further let us notice that we will test the  
CDM halos   at  $R_{opt}$, where the  baryon  infall has not  
significantly  altered the original DM halo {\it velocity 
profile}: $R_{opt}$ is external to the region into  which  most of the  
baryons have collapsed   (Blumenthal et al, 1986). 
Violent dark halos-baryonic matter couplings,  such  as those   proposed by Navarro et al, (1996)
and   Gelato and Sommer-Larsen, (1998) 
 can instead   modify the original halo distribution everywhere;   however, given the very 
heuristic  nature of these processes, it is best  to   first compare the standard-infall   CDM
halos  with the galactic halo, and  then   to  consider the possibly emerging
 discrepancies  in terms of  new
 theoretical scenarios.

The highest possible value for   $\nabla_h^{CDM}$ is $0.5$,
 that is  achieved on the $\sim 10 kpc $  scale  only  for  $c<5$ (see Bullock et al. 1999 and   Navarro,
1998),  i.e.  for   low values of the  concentration parameter, a  property  of 
low-$\Omega$ universes. This  value  is quite inconsistent with the average value  
found in   spiral dark halos, especially if one  considers  that   high resolution 
N-body simulations  converge to a  maximum value of      $\nabla_h^{CDM}=1/4$ (Moore et al, 1998).

Of crucial importance is  also the absence of  a  significant scatter  in the 
 $\nabla_h$ vs. $ log V_{opt}$ relationship. 
 In fact, the CDM  theory  predicts  that,  in   a very  wide region
centered  at  $\sim 10 kpc$ and  certainly  including  $R_{opt}$ 
independently of its  relation with  the virial radius,   galactic    halos
with the same  mass    do not follow   a unique   velocity  curve   but    a 
family of  them. These  can be  described by  a  set of  straight-lines with 
slopes varying between    $-0.1$ and $+0.5$ (e.g. see  fig 6 of  Bullock et
al, 1999). According to CDM the $\nabla_h-log V_{opt}$ plane should be filled
well beyond the tiny strip of Fig  (4).   Taken at its  face value,   the
observational constraint variance(   ${\nabla_h}<0.1 $) could  imply,   within
the   CDM scenario,  that protospiral halos  are co-eval  and have  similar  
merging histories.  A second  possibility  may be that the disk lenght-scale
$R_{opt}$,  in units of virial radius,  is   {\it strongly}   coupled with 
the  structure of the dark matter halo. (e.g., Mo, Mao \& White 1998;
Dalcanton, Spergel \& Summers 1997; van den Bosch et al. 1999, but see also 
Bullock et al, 1999).

  The sizes  of the DM   core radii  turn out to be  very large,   
 at least  $(3-4) R_d$, and    independent of galaxy luminosity  and  
  DM mass  fraction. This may pose a problem  for the suggestion  
 that they  were  formed through some    luminous-dark dynamical coupling
and it  may  call or for   a    primordial
origin for these "warm"    regions  of  constant density embedding   
  the luminous matter or for some dissipative self-regulating" process.

Burkert (1995) and Salucci \& Burkert (2000), from  the analysis of  individual RC's and of the
URC,   have proposed that the 
DM halos around disk systems follow  the   distribution:  $\rho_B(R) \propto (R+R_
B)^{-1} (R^{-2}+R_B^{-2})$,
 i.e.  an NFW profile at large radii which
converges   to a constant value at inner   radii. The core radius,
  $R_B$  is found to increase  from  5 kpc to   
30 kpc,  as  $V_{opt}$ increases form  $75 km/s$ to  $300 km/s$. This implies f
or 
the DM halo: $\nabla_
B \simeq 0.73 $
almost independently   of $V_{opt}$  and  in very good agreement  with the  halo slope
determinations of fig (3).

Then, since the derived density distribution  of galaxy halos is   quite
different from the    theoretical  one out to the outermost velocity data
available (see also Corbelli and Salucci, 1999;  Salucci \& Burkert, 2000), 
we should seriously  begin to consider the possibility that  
cosmological  processes   have cut  the link between the  initial
conditions and the present-day galaxies  properties.

\section {Aknowledgments}

I thank John Miller, Cedric Lacey and Ezio Pignatelli  for very  useful discussions.
and  the anonymous referee for suggestions that have improved the presentation 
of my work.
\noindent 

\vglue 0.5truecm
\ref{Borriello A. and Salucci P., 2000,  submitted to MN, astro-ph}
\ref{Bosma, A. 1981, AJ, 86, 1825}
\ref{Broeils, A.H. 1992, Ph.D. thesis, Groningen University}
\ref{Blumenthal, J.J., Faber, S.M., Flores, R, Primack, J.R., 1986  ApJ, 301,27}
\ref{Bullock, J. S., Kolatt, T. S., Sigad, Y., Somerville, R. S., Kravtsov,
A. V., Klypin, A. A., Primack, J. R., \& Dekel, A. 1999, 
astro-ph/9908159} 
\ref{Burkert, A. 1995, ApJ , 447, L25}
\ref{Burkert, A., \& Silk, J. 1997, ApJ, 488, L55}
\ref{Corbelli E. \& Salucci, P. 1999, MNRAS,   in press, astroph/9909252}  
\ref{Dalcanton, J., Summers,F., Spergel, D., 1997, ApJ 482, 659}
\ref{Fall, S.M., \& Efstathiou, G. 1980, MNRAS, 193, 189}
\ref {Flores, R., \& Primack, J. R. 1994, ApJ , 427, L1}
\ref{Gelato, S. \& Sommer-Larsen, J. , 1999,  MNRAS,  303, 321.}
\ref{Mo, H., Mao, S., White, S.D.M.,  1998, MNRA,  295, 319} 
\ref{Moore, B. 1994, Nature, 370, 629}
\ref {Moore , B., Governato, F., Quinn, T, Stadel.J., Lake, G., 1998, ApJ,  L5}
\ref{Navarro, J. F., Frenk, C. S., \& White, S. D. M. 1996, ApJ, 462, 563}
\ref{Navarro, J.F., Eke V. R.,  Frenk, C.S.F. ,1996,  MNRAS,  283 L72}
\ref{ Navarro J.F.  1998, preprint (astro-ph 9807084)}
\ref{Persic, M., \& Salucci, P. 1988, MNRAS, 234, 131}
\ref{Persic, M., \& Salucci, P. 1990a, ApJ, 355, 44}
\ref{Persic, M., \& Salucci, P. 1990b, MNRAS, 245, 577}
\ref{Persic, M., \& Salucci, P. 1990c, MNRAS, 247, 349}
\ref{Persic, M., \& Salucci, P. 1991, ApJ, 368, 60 }
\ref{Persic, M., \& Salucci, P. 1995, ApJSS, 99, 501 }
\ref {Persic, M., Salucci, P. \& Ashman,  1993, A\&A 279, 343} 
\ref{Persic, M. , Salucci, P.  \& Stel F.  1996, MNRAS, 281, 27 (PSS)}
\ref {Ratnam, C. \& Salucci, P., 2000, New Astron, in press}
\ref{Rhee, M-H, 1997, PhD. Thesis}
\ref {Rubin, V.C., Thonnard, N. \& Ford, W.K., Jr. 1980,
    ApJ, 238,471}
\ref{Salucci P., Burkert,  A, 2000,  ApJ, 312,L27}     
\ref{Salucci P., Persic M, 1997, in  Dark and Visible Matter in Galaxies, ASP 117,1}
\ref{Salucci P., Persic M, 1999a MNRAS. 309, 923 }                           
\ref{Salucci P., Persic M, 1999b  A\&A  351, 442}
\ref{Salucci P., Ratnam, C., Monaco P., Danese, 2000, MNRAS,  in press (astro-ph/9812485)} 
\ref{Salucci, P., Ashman, K.M., \& Persic, M. 1991, ApJ, 379, 89}
\ref{van Albada, T.S., Bahcall, J.N, Begemann, K., Sancisi, R., 1985, ApJ, 295, 305}
\ref{van den Bosch, F. C. 2000, ApJ, 530, 177}

\ref{Verheijen, M., 1997, PhD. Thesis  Thesis, Groningen University}

\vfill\eject
\end{document}